\documentclass[runningheads]{llncs}
\usepackage{cite}
\usepackage{siunitx}
\usepackage{adjustbox}
\usepackage{dingbat}
\usepackage{multirow}

\usepackage[T1]{fontenc}
\def\doi#1{\href{https://doi.org/\detokenize{#1}}{\url{https://doi.org/\detokenize{#1}}}}

\usepackage{graphicx}
\usepackage{pgfplots} 
\pgfplotsset{width=\textwidth, compat=1.6}  

\usepackage{listings}
\usepackage{comment}

\begin{document}
\raggedbottom
\title{Multi-Task Lung Nodule Detection in Chest Radiographs with a Dual Head Network}

\titlerunning{Lung Nodule Detection with Dual Head Network}

% \author{Chen-Han Tsai\inst{1}\orcidID{0000-0002-4665-7783}\and
% Yu-Shao Peng\inst{1}}

\author{Chen-Han Tsai\inst{1}\and
Yu-Shao Peng\inst{1}}
% index{Tsai, Chen-Han}
% index{Peng, Yu-Shao}

\authorrunning{C.H. Tsai et al.}

\institute{HTC DeepQ\\
\email{\{maxwell\_tsai,ys\_peng\}@htc.com}}

% \author{***\inst{1}\and *** \inst{1}}
% %
% \authorrunning{***}

% \institute{***\\
% \email{\{***,***\}@***}}

\maketitle              

\begin{abstract}
% Lung nodules can be an alarming precursor to potential lung cancer. Missed nodule detections during chest radiograph analysis remains a common challenge among thoracic radiologists. In this work, we present a multi-task lung nodule detection algorithm using a Dual-Headed Network (DHN). Unlike from past approaches, our algorithm predicts a global-level label indicating nodule presence along with local-level labels predicting nodule locations using a Dual-Headed Network (DHN). Although seemingly superfluous, we demonstrate that such multi-task formulation in combination with a DHN yields favorable nodule detection performance compared with traditional single-task formulations. In addition, we introduce the notion of dual-headed augmentations tailored for DHN's, and we demonstrate its significance in enhancing nodule detection performance for both global and local predictions. 
Lung nodules can be an alarming precursor to potential lung cancer.
Missed nodule detections during chest radiograph analysis remains a common challenge among thoracic radiologists. 
In this work, we present a multi-task lung nodule detection algorithm for chest radiograph analysis.
Unlike past approaches, our algorithm predicts a global-level label indicating nodule presence along with local-level labels predicting nodule locations using a Dual Head Network (DHN). 
We demonstrate the favorable nodule detection performance that our multi-task formulation yields in comparison to conventional methods. 
In addition, we introduce a novel Dual Head Augmentation (DHA) strategy tailored for DHN, and we demonstrate its significance in further enhancing global and local nodule predictions.

\keywords{Nodule Detection \and Chest Radiograph \and Dual Head Network \and Dual Head Augmentation \and Faster R-CNN}
\end{abstract}

\section{Introduction}
\label{Introduction}
Lung cancer ranks among the top causes of cancer-related deaths worldwide. Pulmonary nodule findings, though typically benign, are an alarming sign for potential lung cancer. Given its simple and inexpensive operating cost, chest radiography (i.e., x-rays) is the most widely adopted chest imaging solution available. However, one concern during radiograph analysis is the proportion of nodules thoracic radiologists often miss due to the nature of the imaging modality \cite{missed_lung_cancer, Tack2019}. A chest radiograph is a 2D projection of a patients’ chest. Thus, nodules appear less visible when occluded by other organs (e.g., rib cages) or foreign bodies (e.g., CVADs). With the rising workload already posing a challenge for thoracic radiologists, assistive tools to reduce missed nodules during chest radiograph analysis are gaining significant clinical relevance \cite{Nam2019, Hwang2020, Cha2019}. 

To identify pulmonary nodules on chest radiographs, several works propose potential solutions. Some \cite{Wang2017, Rajpurkar2017, Ausawalaithong2019} focus on image-level prediction indicating nodule presence per scan (we refer to as {\em global} methods). 
Others~\cite{Li2018, Mendoza2020, Kim2019, Schultheiss2020} study patch-level prediction exploring nodule detection with local bounding box information for each nodule (we refer to as {\em local} methods).
Although both local and global methods offer information regarding nodule presence in a given chest radiograph, adopting just a single method alone can be undesirable. Local methods offer the benefit of pinpointing each nodule, but the adopted labeling criterion can be highly subjective and prone to inconsistency \cite{Larici2017, Busby2018}. Global methods alleviate this issue by predicting a single label indicating nodule presence, but further effort is required by the examiner to locate these nodules.

Li et al.~\cite{harvardthoracic} and Pesce et al.~\cite{Pesce2019} attempted to address this disparity.
The former formulated a multiple instance learning (MIL) model to classify nodule presence using a grid of patches across the input scan. The image-level label is then computed using the joint probability across patch predictions. Despite its attempt in combining local and global predictions, such MIL model is not translation equivariant by design, causing the predicted global label to be highly dependent on the nodule location innate to the scan.
The latter proposed CONAF, a network composed of a shared backbone between its classification and localization head. The classification head outputs a global label indicating nodule presence, and its localization head outputs a downsized score map. Considering that most nodule sizes are relatively small with respect to the scan size, the low resolution scoremap limits the localizer head from explicit localization purposes. 

\begin{comment}
    A study by Li et al.~\cite{harvardthoracic} attempted to address this disparity by formulating a multiple instance learning (MIL) model. The proposed model predicts a grid of patches across the input scan. Each patch outputs a value approaching $1$ if a nodule exists in that patch and $0$ otherwise. The image-level label is then computed using the joint probability across patch predictions. Despite its attempt in combining local and global predictions, such MIL model is not translation equivariant by design, causing the predicted global label to be highly dependent on the nodule location innate to the scan.
    
    The CONAF architecture~\cite{Pesce2019} is a CNN architecture that exploits local labels with soft attention feedback. Its architecture consists of a shared backbone between its classification and localization heads. CONAF's classification head outputs a global label indicating nodule presence, and its localization head outputs a downsized scoremap (similar to the grid structure patches in \cite{harvardthoracic} but with a 2D Gaussian kernel). Considering that most nodule sizes are relatively small with respect to the scan size, the low resolution scoremap limits the localization head from explicit localization purposes. 
\end{comment}

In this work, we present a novel multi-task lung nodule detection algorithm using a Dual Head Network (DHN) and an accompanying Dual Head Augmentation (DHA) training strategy. Our multi-task objective is similar to \cite{multitask}, but instead of using a RetinaNet\cite{Lin_2017_ICCV}, we adopt a modified Faster-RCNN\cite{fasterrcnn} architecture customized for nodule detection. Considering the properties of pulmonary nodule, we propose to use deformable convolutions~\cite{deformable} and the gIOU loss~\cite{giou_loss}. In addition, we incorporate our novel DHA strategy during DHN training, and we demonstrate its importance in further enhancing global and local nodule predictions.
The remaining of the paper is organized as follows. We first illustrate the preliminaries in Section \ref{Preliminaries}. The proposed methods will be detailed in Section \ref{Methods}, followed by experimental results in Section \ref{Experiments}. Conclusion will be given in Section \ref{Conclusion}.

\begin{comment}
    In this work, we present a novel multi-task lung nodule detection algorithm using a Dual-Headed Network (DHN). Our multi-task objective is similar to \cite{multitask}, but instead of using a RetinaNet, we adopt a modified Faster-RCNN\cite{fasterrcnn} architecture customized for nodule detection. In addition, we propose a novel Dual-Headed augmentation (DHA) strategy tailored for DHN's, and we demonstrate its importance in further enhancing global and local nodule predictions. The remaining of the paper is organized as follows. The proposed methods will be detailed in Section \ref{Methods}, followed by experimental results in Section \ref{Experiments}. Conclusion and future work will be give in Section \ref{Conclusion}.
\end{comment}

\section{Preliminaries}
\label{Preliminaries}
The Faster-RCNN \cite{fasterrcnn} is a two-stage network originally designed for object detection. The first stage of the network is a feature extractor (i.e. VGG-16\cite{vgg}), and the second stage consists of an RPN, ROIPool, and ROIHead module. For a given input image, the feature extractor outputs a feature representation of that image, and this representation is fed to the second stage. The RPN generates bounding box proposals around regions that contain potential non-background objects. Crops for each proposed region are then taken from the extracted feature representation, and they are resized to a fixed dimension using ROIPool. The fixed-size feature crops are then independently classified using the ROIHead, and an updated bounding-box is predicted for each crop.

Modern implementations of the Faster-RCNN often include two modifications to the original design that enhances detection performance. The first is the attachment of the Feature Pyramid Network (FPN) \cite{fpn} behind the first stage feature extractor. FPN allows cross-level information flow between multi-resolution representations which is beneficial for detecting small objects. The second is the replacement of ROIPool with Multi-Scale ROIAlign \cite{maskrcnn, fpn} to avoid quantization while cropping the multi-resolution features. 

% Modern implementations of Faster-RCNN have been modified in two ways. First, since the introduction of the Feature Pyramid Network (FPN) \cite{fpn}, modern implementations have attached an additional FPN behind the first stage feature extractor. FPN allows feature representations to be aggregated and extracted at multiple resolutions. Second, modern implementations have also replaced ROIPool with Multi-Scale ROIAlign \cite{maskrcnn, fpn} to avoid quantization while cropping the multi-resolution features. 

During training, the RPN generates an objectness loss $\ell_{obj}$ from the background foreground classification and a regression loss $\ell_{reg}$ from the distance computed between the proposed regions with their matched ground truth bounding boxes. The ROIHead generates a classification loss $\ell_{cls}$ during feature crop classification and a bounding box loss $\ell_{bbox}$ from the distance computed between the updated bounding boxes and their matched ground truth bounding boxes. The default setup utilizs a Smooth L1 Loss to train $\ell_{reg}, \ell_{bbox}$ and the Cross Entropy Loss to train $\ell_{obj}, \ell_{cls}$. The final loss function $L_{local}$ is formulated in Equation \ref{faster_rcnn_loss}
\begin{equation}
    \label{faster_rcnn_loss}
    % \vspace{-2mm}
    L_{local} = \ell_{obj} + \ell_{reg} + \ell_{cls} + \ell_{bbox}. 
    % \vspace{-5mm}
\end{equation}

\begin{figure}[t]
    \centering
    \includegraphics[width=\textwidth]{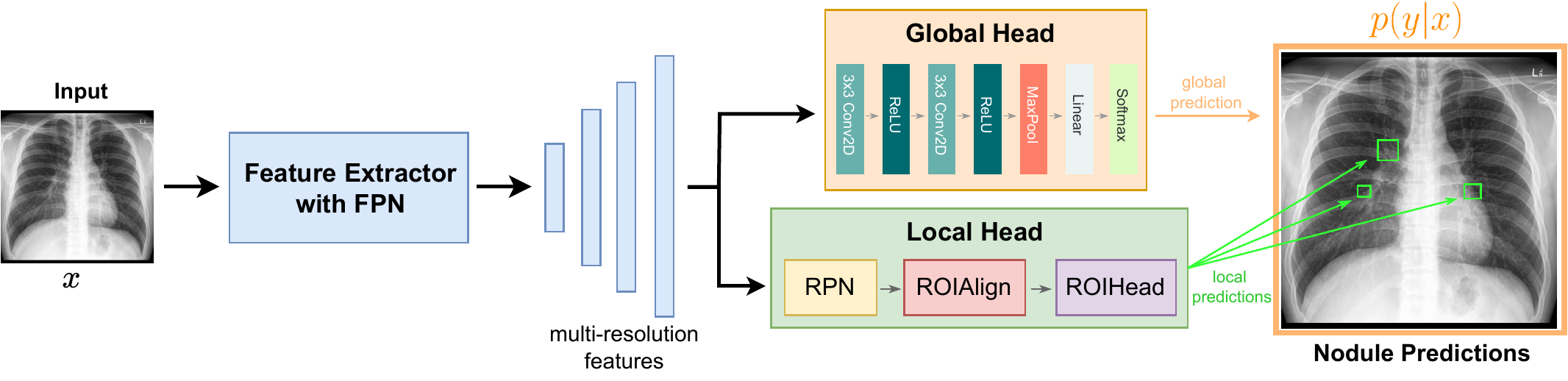}
    % \vspace{-4mm}
    \caption{An illustration of the DHN architecture during inference. The input $x$ is fed through a feature extractor with FPN to obtain a set of multi-resolution features. The \textit{global head} predicts the global label indicating nodule presence in the scan, and the \textit{local head} predicts local bounding boxes indicating nodule locations. }
    % \vspace{-4mm}
    \label{fig:DHN_Architecture}
\end{figure}

\section{Methods}
\label{Methods}
In this section, we present our multi-task Dual Head Network (DHN) architecture and the accompanying Dual Head Augmentation (DHA) strategy. The DHN takes advantage of the two-stage structure seen in the Faster-RCNN by adding an additional network in parallel to the original second stage. The DHA strategy is applied during training, and it utilizes the DHN's dual head design in improving nodule detection performance. The specifics are detailed in the following sections. 

\subsection{Dual Head Network}

\label{dhn_architecture}
The DHN architecture is designed in a two-stage approach (see Figure \ref{fig:DHN_Architecture}). The first stage is a feature extractor with FPN, and the second stage consists of two parallel networks that we refer to as the \textit{global head} and the \textit{local head}. 
For a given scan, 
the feature extractor first extracts its respective representation. Then,
the \textit{global head} predicts a binary label indicating nodule presence, and the \textit{local head} predicts bounding boxes around each detected nodule.
%We highlight the details below.

\begin{figure}[t]
    \centering
    \includegraphics[width=.85\textwidth]{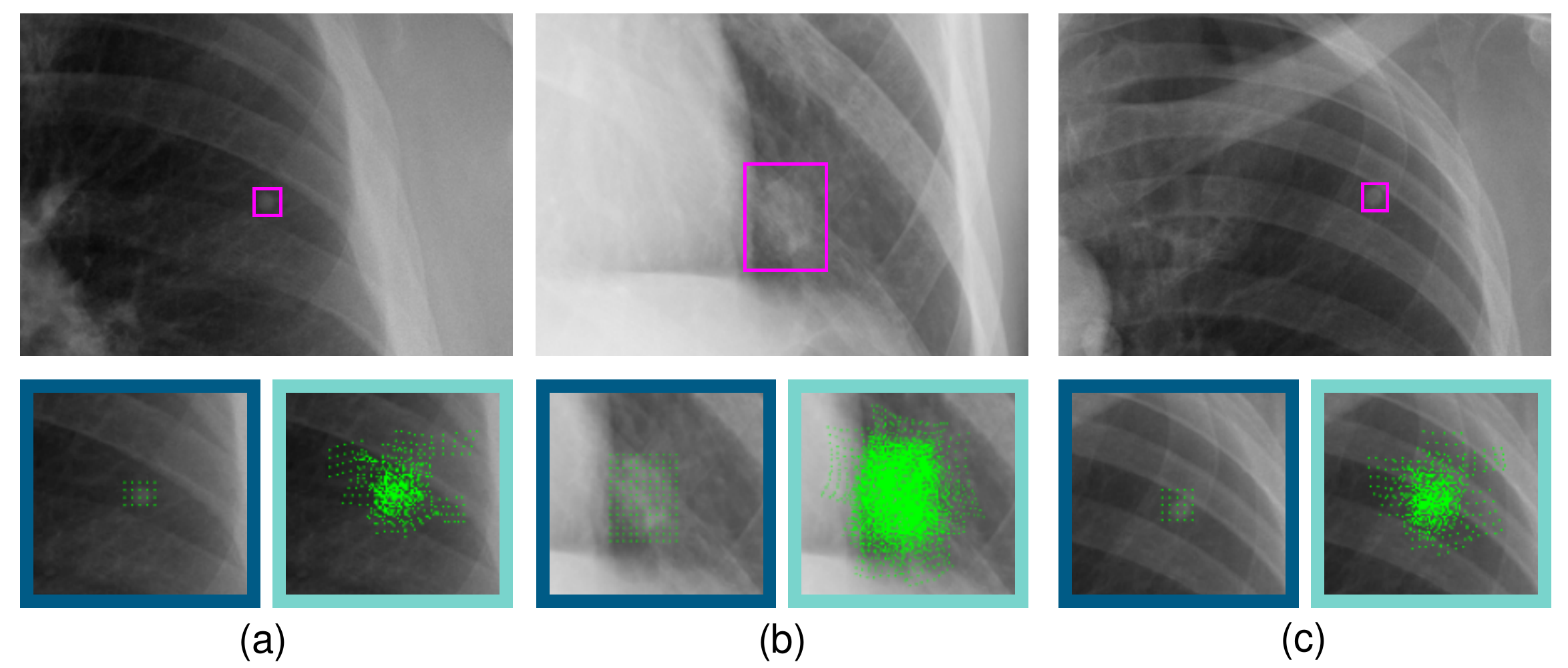}
    % \vspace{-3mm}
    \caption{Comparison between deformable (turquoise) versus standard convolution (dark blue) applied on three sample cases (a-c) using the 3rd layer of a trained ResNet-18 feature extractor. The receptive fields are fixed if regular convolution is applied, whereas deformable convolutions allow dynamic focus on the regions of interest.}
    % \vspace{-4mm}
    \label{fig:deformable_example}
\end{figure}

The feature extractor of our DHN is a modified ResNet-18\cite{resnet}. We conduct a series of experiments comparing various CNN architectures from the ResNet family, and we observe slightly better performance as model size increased. However, the training time increases significantly for larger models. Hence, we select the smallest model from the ResNet family, the ResNet-18, to serve as a lower bound for DHN performance throughout our experiments. We modify the ResNet-18 by replacing standard convolutions with deformable convolutions \cite{deformable} in the final three layers. Applying deformable convolutions allows more dynamic focus on particular regions in the image where the nodules size might be small (see Figure \ref{fig:deformable_example}). For a given scan $x$, we take the intermediate representations the ResNet-18 extracts, and we pass them to the FPN to obtain set of multi-resolution representations. This set of multi-resolution representations is fed to the \textit{global} and \textit{local heads} for further processing.

The \textit{global head}'s primary purpose is to classify whether nodules are present in a given scan. Thus, we select the representation with the largest receptive field \cite{Wenjie_Receptive_Field_NIPS16} as input. As shown in Figure~\ref{DHA_fig}, this representation is passed through two consecutive 2D convolutions and ReLUs before being max-pooled into a single vector. 
Then, a linear layer and a softmax layer are applied to obtain the probability indicating nodule presence in the scan. 
%This vector is affine transformed to a length of two, then softmax is applied to obtain the probability $p(y|x)$ indicating nodule presence in the scan. 
For a set of $N$ labeled scans $\{(x_i, y_i)\}_{i=1}^N$ where $x_i\in\mathbf{R}^{h\times w}$ is the $i^{th}$ scan with resolution $h\times w$ (we set $h,w=512$) and $y_i\in\{0,1\}$ is the corresponding global label, we formulate the \textit{global head loss} $L_g$ (see Equation \ref{global_label_loss}) using a weighted cross-entropy:
%\textcolor{red}{($\alpha_1, \alpha_2$ are discussed in Section \ref{preprocessing})}.
% \vspace{mm}
\begin{equation}
\label{global_label_loss}
    L_g = -\sum_i^N \alpha_{1}y_i\log{(p(y_i|x_i))} + \alpha_{2}(1-y_i)\log{(1-p(y_i|x_i))},
% \vspace{-1mm}
\end{equation}
where $\alpha_1$ and $\alpha_2$ are two hyperparameters.

The \textit{local head} serves as a nodule detector, and we adopt the second-stage design as specified in Section \ref{Preliminaries}. During training, however, we replace the Smooth L1 loss in the RPN and the ROIHead with the gIOU loss \cite{giou_loss} since the gIOU's scale-invariant property is beneficial during small scale bounding box optimization. The \textit{local head loss} $L_l$ follows the formulation in Equation \ref{faster_rcnn_loss}. We propose an end-to-end optimization method to consider the \textit{local head loss} and the \textit{global head loss} simultaneously. The final multi-task loss $L$ (see Equation \ref{multitask_loss}) is a weighted sum of the two losses, i.e.,
\begin{equation}
\label{multitask_loss}
    L = \lambda_1 L_g + \lambda_2 L_l.
\end{equation}

\subsection{Dual Head Augmentation}
\label{dual_head_aug}
Data augmentation is a well-known technique that increases diversity in the training data to improve model generalization. In classical image classification or object detection tasks, one augmentation strategy (i.e., a pre-defined set of stochastic transforms) is applied per training image during the forward-pass. 
However, training with just one augmentation strategy per image for a dual head architecture can easily lead to one head being particularly optimized while the other head performs mediocrely.

\begin{figure}[t]
    \centering
    \includegraphics[width=\textwidth]{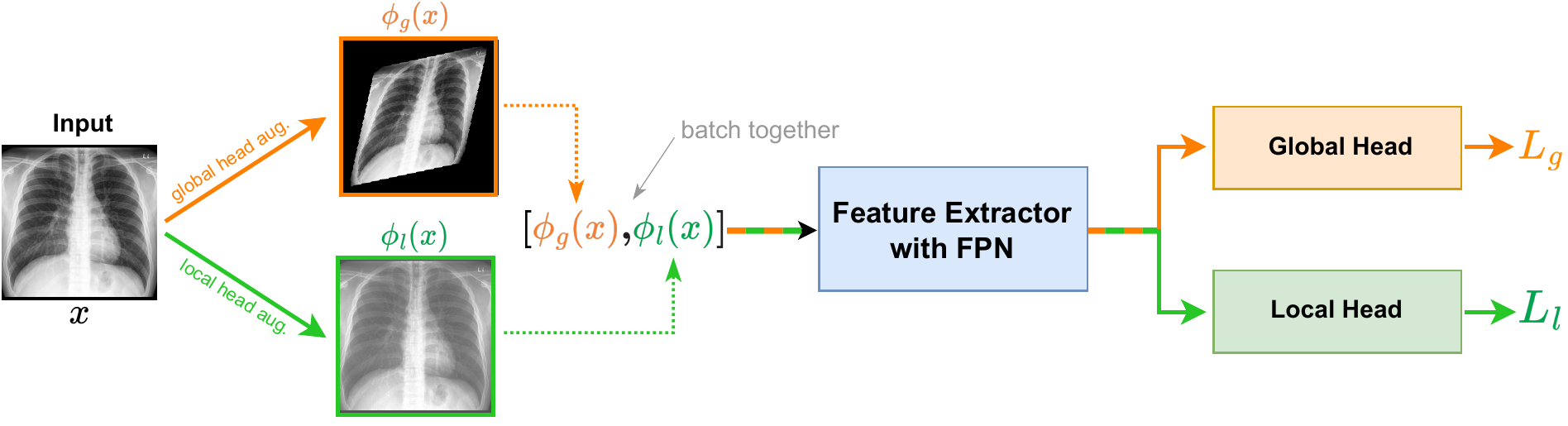}
    % \vspace{-2mm}
    \caption{An illustration of DHA. Two augmented images $\phi_g(x)$ and $\phi_l(x)$ of $x$ are generated. They are batched together and passed into the feature extractor with FPN. The output features are then split according to their corresponding head. Individual head losses $L_g$ and $L_l$ are summed during training. Notice that $\phi_g(x)$ only updates the \textit{global head}, and $\phi_l(x)$ only updates the \textit{local head} during back-propagation.}
    % \vspace{-3mm}
    \label{DHA_fig}
\end{figure}

To fully optimize each head to their specific objectives, we propose a novel data augmentation strategy called the Dual Head Augmentation (DHA). DHA takes advantage of the DHN's dual head structure by applying an augmentation strategy for each head. As shown in Figure \ref{DHA_fig}, we designate an augmentation function $\phi_g$ for the \textit{global head} and $\phi_l$ for the \textit{local head}. 
Given an input scan $x$, the augmented images $\phi_g(x)$ and $\phi_l(x)$ are generated and batched together. This batch is fed into the feature extractor with FPN, and we obtain the multi-resolution representations. We split the batch, and we feed the representations corresponding to $\phi_g(x)$ and $\phi_l(x)$ to the \textit{global} and \textit{local heads} respectively to optimize each head.

We select a set of image transformations\footnote{The complete list of transformations are detailed in the supplementary materials.} that involves illumination transforms (e.g., histogram equalization, random brightness, etc.) and geometric transforms (e.g., horizontal flip, rotation, etc.).
%, and the complete list is provided in the supplementary materials. 
In this work, we consider two types of augmentation strategies that exhibit favorable single-head performance, and we define the augmentation strategies by their sampling method. The first sampling method is binomial sampling. Formally, for a given probability $p$ and a set of transforms $\Theta = \{ \theta_j \}_{j=1}^M$, binomial sampling selects transform $\theta_j$ for $j \in \{1, ..., M \}$ with probability $p$, and we refer to the corresponding augmentation strategy as $\phi^{\texttt{bin}}(x;p, \Theta)$. %(we set $p=0.9$). 
The second sampling method is uniform sampling. Specifically, uniform sampling selects a transform from $\Theta$ with probability $1/M$, and we refer to this augmentation strategy as $\phi^{\texttt{uni}}(x;\Theta)$. During each training iteration, the augmentation strategy samples from $\Theta$ using one of the above sampling methods and applies the selected transforms.

\section{Experiments}
\label{Experiments}
In this section, we evaluate the nodule detection performance of a DHN in comparison to other notable approaches. In addition, we perform a series of experiments to compare the influence different augmentation strategies play towards enhancing global and local predictions. 
Classification performance is evaluated using the \textit{global head} prediction, and localization performance\footnote{FROC and AFROC are computed with an Intersection over Union (IOU) threshold of 0.4, and the FROC-AUC is computed with a False Positive Per-Image up to 1.} 
is evaluated with \textit{local head} predictions. We report on the test set mean and standard deviation of the top 8 performing
%\footnote{Global and local metrics were unified by taking the mean of the global head's ROC-AUC and the local head's mAP metric.} 
validation checkpoints throughout each experiment. 
Each model is trained with SGD and momentum for 70 epochs ($\sim$3 days on an NVIDIA P100 16GB GPU) with a step size of 5e-5 and momentum of 0.975 at a batch size of 2. 
Loss weights $\alpha_1$, $\alpha_2$ are set to 0.69 and 1.76 while $\lambda_1$, $\lambda_2$ are set to 0.35 and 2.5.
%We set parameters $\alpha_1=0.55$, $\alpha_2=4.9$ and $\lambda_1=0.35$, $\lambda_2=2.5$.

\label{dual_head_comparison}
%\vspace{-4mm}
\begin{table}[t]
\begin{adjustbox}{width=\textwidth}
\begin{tabular}{l|cc|ccc}
\hline
\hline
\multirow{2}{*}{Model} & \multicolumn{2}{c|}{Classification Metrics} & \multicolumn{3}{c}{Localization Metrics} \\ 
& \textbf{ROC-AUC}                  & \textbf{PR-AUC}                & \textbf{FROC-AUC}               & \textbf{AFROC-AUC}              & \textbf{mAP}                   \\ \hline
FCOS \cite{fcos} &     -       &     -      &   $0.540\pm0.054$        &     $0.564\pm0.024$        &    $0.166\pm0.009$         \\
RetinaNet \cite{Lin_2017_ICCV} &     -       &     -      &   $0.576\pm0.009$        &     $0.507\pm0.023$        &    $0.166\pm0.003$         \\
DenseNet-121 \cite{Huang2017} &     $0.856\pm0.006$       &     $0.612\pm0.006$         &   -        &     -        &    -         \\
CONAF \cite{Pesce2019}  &     $0.800\pm0.004$       &     $0.450\pm0.007$         &   -        &     -        &    -         \\
Global Head only  &     $0.809\pm0.010$       &     $0.524\pm0.014$         &   -        &     -        &    -         \\
Local Head only  &     -       &     -         &   $0.618\pm0.006$        &     $0.606\pm0.029$        &    $0.184\pm0.008$         \\
DHN & $\textbf{0.873}\pm\textbf{0.004}$   & $\textbf{0.654}\pm\textbf{0.016}$  & $\textbf{0.628}\pm\textbf{0.011}$ & $\textbf{0.626}\pm\textbf{0.033}$ & $\textbf{0.188}\pm\textbf{0.009}$ \\ \hline \hline

\end{tabular}
\end{adjustbox}
% \vspace{.25pt}
\caption{\label{tab:dual_head_comparison} Performance comparison between the DHN and alternate methods trained without data augmentation.}
% \vspace{-8mm}
\end{table}

\subsection{Dataset}
The dataset we evaluate in this study is a subset of the National Institute of Health (NIH) Chest X-Ray dataset \cite{Wang2017}. Each case in our dataset was labeled with bounding box annotations for each nodule by three thoracic radiologists. Prior to labeling, each radiologist had to pass an assessment test requiring them to correctly identify over $80\%$ of the cases from a held-out test set with potential lung nodules. The radiologists were then asked to independently label each case in our dataset. Labels were aggregated upon completion, and the radiologists reviewed each case to reach a consensus on which labels to keep. 
%A final senior radiologist then reviewed and modified, if necessary, the annotations the three radiologists had agreed upon. Scans that remain unmodified following the senior radiologist's review were added to our dataset. In total, our dataset consists of 21,189 scans. The scans were then split into training, validation, and test sets following an 80:10:10 ratio based on patient identifiers.
A final senior radiologist then reviewed and control the quality of annotations. In total, we randomly sample 26000 scans from NIH dataset and 21,189 qualified scans are added to our dataset. These scans are split into training, validation, and test sets following the ratio of $80:10:10$ based on patient identifiers.

% \subsection{Model Optimization}
% We utilize Tune \cite{liaw2018tune} to select hyperparameter throughout our experiments. The parameters $\lambda_1$ and $\lambda_2$ of the multi-task loss $L$ are set to $0.35$ and $2.5$ respectively. SGD with momentum yielded favorable results with a learning rate of 2.1e-4, momentum at 0.975, weight decay of 4e-5, and a batch size of 2. Binomial sampling probability $p$ was set at $0.9$. The weights $\alpha_1$ and $\alpha_2$ of the \textit{global head loss} $L_g$ are set to $0.55$ and $0.49$. We utilize these hyperparameters to train each model for 70 epochs which takes approximately 2 days using an NVIDIA-P100 16GB GPU. The feature extractor is a ResNet-18 pretrained on ImageNet, and the first three layers are fixed during training.
% The reported localization metrics are computed with an Intersection over Union (IOU) threshold of 0.4, and the FROC-AUC is computed with a False Positive Per-Image up to 1. 

\subsection{Dual Head Network Analysis}
\label{exp:dhn_analysis}

For our first set of experiments, we evaluate the performance of a DHN in comparison to several \textit{global} and \textit{local} methods \cite{fcos, Lin_2017_ICCV, Huang2017}. We also compare our DHN with a notable dual-head approach CONAF \cite{Pesce2019}. Although CONAF's localization head generates a heatmap-like mask for nodule localization, the mask is too coarse to derive precise bounding box coordinates from. Hence, we only evaluate the classifier head's prediction in our comparison.
For fair comparison, we do not apply data augmentation to the training strategy. We also analyze the single head performance of the DHN. Specifically, we consider the case where we train only the \textit{global head} and the case where we train only the \textit{local head}.

As shown in Table \ref{tab:dual_head_comparison}, the DHN architecture yields favorable improvements in both classification and localization metrics.
%In this set of experiments, we evaluate the performance of a DHN in comparison to a notable approach CONAF \cite{Pesce2019}.  \textcolor{red}{We also include several \textit{global} and \textit{local} methods prior works were based upon as part of our experiment\cite{fcos, Lin_2017_ICCV, Huang2017}.} For fair comparison, we do not apply data augmentation to the training strategy. We also analyze the single head performance of the DHN. Specifically, we consider the case where we train only the \textit{global head} and the case where we train only the \textit{local head}. As shown in Table \ref{tab:dual_head_comparison}, the DHN architecture yields favorable improvements in both classification and localization metrics. 
We believe this improvement is that our DHN extracts more informative representations due to cross-task supervision.
%We believe this improvement is the result of a more informative feature representations a DHN extracts due to cross-task supervision.
Specifically, training with local labels using the \textit{local head} encourages the feature extractor to extract more meaningful representations that highlight local findings. Since these representations are shared with the \textit{global head}, classification can benefit from the available local information the representations possess. Conversely, global labels are trained with high level features that have the widest receptive field. This encourages the feature extractor to leverage potential ancillary information beneficial for global prediction which may also assist local predictions. As our experiments demonstrate, we observe complementing nodule detection performance from the two heads when they are jointly trained.

%\textcolor{blue}{In comparison to CONAF, we evaluate only the predictions from the classifier head. Although CONAF's localization head generates a heatmap-like mask for nodule localization, the mask is too coarse to derive precise bounding box coordinates from. Hence, we only evaluate the classifier head's prediction in our comparison.}

% {When evaluating CONAF, we consider only the classification metrics since the classifier head prediction was the only output that can be directly compared. Although CONAF has a dual-head architecture (i.e., the one for classification and the second for localization), extracting precise bounding box coordinates from the coarse localization map generated by the second head is not a trivial task. Furthermore, a true positive prediction in our localization evaluation requires an IOU > 0.4 between the ground truth and predicted boxes. This requirement is much stricter than the criterion adopted by \cite{Pesce2019} in which a predicted mask is labeled true positive when it overlaps the ground truth area by > 25\%. Thus, we consider only the classification head prediction of CONAF as part of our evaluation.}

\subsection{Dual Head Augmentation Evaluation}
\label{exp_dual_head_aug}
In this subsection, we are interested in analyzing the effects different augmentation strategies impose on the DHN's global and local predictions. As mentioned in Section \ref{dual_head_aug}, we consider both binomial and uniform augmentation strategies $\phi^{\texttt{bin}}$ and $\phi^{\texttt{uni}}$. We also include an identity transform (no augmentation) strategy denoted $\phi^{\texttt{id}}$ as a baseline reference. 

\begin{table}[]
% \vspace{-4mm}
\begin{adjustbox}{width=\textwidth}
\begin{tabular}{ll|cc|ccc}
\hline
\hline
\multicolumn{2}{c|}{DHA}         &  \multicolumn{2}{c|}{Classification Metrics} & \multicolumn{3}{c}{Localization Metrics}                            \\ 

\textbf{Global} & \textbf{Local}  & \textbf{ROC-AUC}     & \textbf{PR-AUC}      & \textbf{FROC-AUC}    & \textbf{AFROC-AUC}   & \textbf{mAP}         \\ \hline
$\phi^{\texttt{id}}$                   & $\phi^{\texttt{id}}$                   & $0.873\pm0.004$          & $0.654\pm0.016$            & $0.628\pm0.011$          & $0.626\pm0.033$            & $0.188\pm0.009$          \\
$\phi^{\texttt{bin}}$                  & $\phi^{\texttt{id}}$  & $\textbf{0.879}\pm\textbf{0.007}$  & $\textbf{0.675}\pm\textbf{0.017}$ & $0.624\pm0.012$           & $0.575\pm0.037$          & $0.184\pm0.006$          \\
$\phi^{\texttt{uni}}$                  &  $\phi^{\texttt{id}}$             & $0.877\pm0.004$           & $0.673\pm0.012$          & $0.627\pm0.009$          & $0.597\pm0.057$          & $0.182\pm0.06$           \\
$\phi^{\texttt{id}}$  & $\phi^{\texttt{bin}}$                  & $0.858\pm0.007$          & $0.631\pm0.011$          & $0.656\pm0.009$          & $\textbf{0.687}\pm\textbf{0.014}$ & $0.182\pm0.007$          \\
$\phi^{\texttt{id}}$          & $\phi^{\texttt{uni}}$                  & $0.877\pm0.003$          & $0.643\pm0.009$          & $\textbf{0.658}\pm\textbf{0.017}$  & $0.663\pm0.015$          & $\textbf{0.200}\pm\textbf{0.005}$  \\ \hline
$\phi^{\texttt{bin}}$                  &  -  & $\textbf{0.849}\pm\textbf{0.004}$ & $\textbf{0.628}\pm\textbf{0.007}$ & -                    & -                    & -                    \\
$\phi^{\texttt{uni}}$                  &        -              & $0.834\pm0.008$          & $0.585\pm0.019$          & -                    & -                    & -                    \\
-  & $\phi^{\texttt{bin}}$                  & -                    & -                    & $0.637\pm0.032$ & $\textbf{0.686}\pm\textbf{0.028}$  & $0.177\pm0.024$          \\
-  & $\phi^{\texttt{uni}}$                  & -                    & -                    & $\textbf{0.651}\pm\textbf{0.021}$          & $0.662\pm0.015$           & $\textbf{0.204}\pm\textbf{0.008}$ \\ \hline \hline

\end{tabular}
\end{adjustbox}
% \vspace{.25pt}
\caption{\label{tab:dual_head_aug_comp1} Comparison of single head DHA strategies ($\phi^{\texttt{id}}$ on one of the heads). First five rows are the results of joint training with single head DHA. The bottom four rows are the results when only a single head is trained with the specified augmentation strategy.}
% \vspace{-8mm}
\end{table}

We first analyze the performance of a DHN when only one augmentation strategy is applied on one of the two heads. From the results shown in Table \ref{tab:dual_head_aug_comp1}, we can see that better classification performance is observed when the \textit{global head} uses the strategy $\phi^{\texttt{bin}}$. Similarly, we observe slightly better localization performance when the \textit{local head} adopts the strategy $\phi^{\texttt{uni}}$. We can also observe noticeable improvements with dual head training versus single head training. When only one augmentation strategy is employed on a single DHN head, its performance still surpasses that of a single head network that utilizes the same augmentation strategy. This observation reflects the results we perceive in Section \ref{exp:dhn_analysis}. Thus, even when just one augmentation strategy is applied on a single DHN head, we can expect favorable performance from DHN's over single head networks.

\begin{table}[]
% \vspace{-4mm}
\begin{adjustbox}{width=\textwidth}
\begin{tabular}{cc|cc|ccc}
\hline \hline
\multicolumn{2}{c|}{DHA} & \multicolumn{2}{c|}{Classification Metrics}  & \multicolumn{3}{c}{Localization Metrics}                           \\
\textbf{Global}      & \textbf{Local}   & \textbf{ROC-AUC}     & \textbf{PR-AUC}      & \textbf{FROC-AUC}    & \textbf{AFROC-AUC}   & \textbf{mAP}         \\ \hline
$\phi^{\texttt{bin}}$         & $\phi^{\texttt{bin}}$        & $0.882\pm0.003$          & $0.674\pm0.008$          & $0.668\pm0.008$          & $\textbf{0.707}\pm\textbf{0.013}$ & $0.187\pm0.006$          \\
$\phi^{\texttt{uni}}$         & $\phi^{\texttt{uni}}$        & $0.881\pm0.009$           & $0.675\pm0.015$           & $0.664\pm0.008$           & $0.702\pm0.009$          & $0.159\pm0.012$ \\
$\phi^{\texttt{bin}}$         & $\phi^{\texttt{uni}}$        & $\textbf{0.903}\pm\textbf{0.003}$ & $\textbf{0.702}\pm\textbf{0.003}$ & $\textbf{0.708}\pm\textbf{0.005}$ & $0.705\pm0.008$          & $\textbf{0.245}\pm\textbf{0.005}$ \\
$\phi^{\texttt{uni}}$         & $\phi^{\texttt{bin}}$        & $0.878\pm0.003$          & $0.667\pm0.010$          & $0.658\pm0.011$           & $0.676\pm0.011$          & $0.202\pm0.005$          \\ \hline \hline
\end{tabular}
\end{adjustbox}
% \vspace{.25pt}
\caption{\label{tab:dual_head_aug_comp2} Comparison of dual head DHA strategies (i.e., $\phi^{\texttt{bin}}$ and $\phi^{\texttt{uni}}$ applied on the global and local heads).}
% \vspace{-10mm}
\end{table}

With the observed characteristics for the \textit{global} and \textit{local head}, we propose a hypothesis that heavier augmentations $\phi^{\texttt{bin}}$ are more suitable for the \textit{global head} but not necessarily the \textit{local head}. 
We believe that the \textit{global head} favors heavier augmentations to increase its robustness against image-level distortions. In contrary, the \textit{local head} seems to prefer lighter augmentations as heavy augmentations might have imposed an overwhelming amount of distortions to region specific features.
We verify our hypothesis with an experiment shown in Table \ref{tab:dual_head_aug_comp2}. The results verify our hypothesis that the DHN obtains optimal performance when the DHA strategy applies $\phi^{\texttt{bin}}$ on the \textit{global head} and $\phi^{\texttt{uni}}$ on the \textit{local head}.

\begin{comment}
    we would like to understand how a DHN behaves when the \textit{global head} and \textit{local head} adopt $\phi^{\texttt{bin}}$ and $\phi^{\texttt{uni}}$ together. 
    We verify our hypothesis with the results shown in Table \ref{tab:dual_head_aug_comp2}.
    %With the observed characteristics for the global and local head, we would like to understand how a DHN behaves when the \textit{global head} and \textit{local head} adopt $\phi^{\texttt{bin}}$ and $\phi^{\texttt{uni}}$ together. 
    As intended, the DHN performs optimally with our hypothesized configuration in comparison to the remaining DHA strategies. 
    A possible reason for this phenomenon is that the \textit{global head} favors heavier augmentations to increase its robustness against image-level distortions. In contrary, the \textit{local head} seems to prefer lighter augmentations as heavy augmentations might have imposed an overwhelming amount of distortions to region specific features. This is the best performing DHA strategy we have obtained to date. 
\end{comment}

\section{Conclusion}
\label{Conclusion}
In this work, we present a multi-task lung nodule detection algorithm using a DHN. Our DHN architecture features a \textit{global head} and a \textit{local head} that performs lung nodule detection on the global and local level simultaneously. In addition, we introduce a novel DHA strategy that leverages the dual head design of the DHN to enhance global and local nodule detection performance while training. Throughout our experiments, we demonstrate the performance gain our DHN yields in comparison to conventional single head networks in both classification and localization abilities. 
% Furthermore, we identified the appropriate augmentation strategy for each head, and our DHN obtained a performance otherwise not attainable if regular single head augmentation strategies are employed. 
Furthermore, we identified the DHA strategy that applies the appropriate augmentations for each head. Together with the optimal DHA strategy, our DHN obtained a performance otherwise not attainable if regular single head augmentation strategies are employed.

\subsubsection{Acknowledgements} 
We would like to thank Che-Han Chang and the anonymous reviewers for their valuable suggestions. We also thank the members: Chun-Nan Chou, Fu-Chieh Chang, Yu-Quan Zhang, and Hao-Jen Wang for their support in collecting annotated data, and Yi-Hsiang Chin for his efforts in conducting experiments.

%Future works may include adding attention between the \textit{global} and \textit{local heads} so that the predictions can be cross evaluated before being output. Additional research is also needed to automate the selection of the sampling strategy relevant to each head. With the findings thus far, we believe our proposed DHN architecture along with the DHA strategy may serve as a firm basis for future research involving lung nodule detection. 

\bibliographystyle{splncs04}
\bibliography{paper613}

\end{document}